# Machine Learning for Offensive Security: Sandbox Classification Using Decision Trees and Artificial Neural Networks

by Will Pearce[1], Nick Landers[1], and Nancy Fulda[2]

[1] SilentBreak Security, Lehi, UT 84043, USA,
{will, nick}@silentbreaksecurity.com,
[2] Brigham Young University, Provo, UT 84602, USA,
nfulda@cs.byu.edu

**Abstract.** The merits of machine learning in information security have primarily focused on bolstering defenses. However, machine learning (ML) techniques are not reserved for organizations with deep pockets and massive data repositories; the democratization of ML has lead to a rise in the number of security teams using ML to support offensive operations. The research presented here will explore two models that our team has used to solve a single offensive task, detecting a sandbox. Using process list data gathered with phishing emails, we will demonstrate the use of Decision Trees and Artificial Neural Networks to successfully classify sandboxes, thereby avoiding unsafe execution. This paper aims to give unique insight into how a real offensive team is using machine learning to support offensive operations.

**Keywords:** neural networks, malware, detection, offensive, machine learning, information security

## 1 Introduction

The composite set of problems an offensive team needs to solve in order to gain and keep access to a network is quite complex, especially when there are one or more defensive products at each phase of an attack. At a very high level, the process [8] is as follows:

1. **External Reconnaissance** – Gathering emails, footprinting external infrastructure.

2. **Initial Access** – Exploiting a technical vulnerability, or landing a phish.

3. **Foothold Stabilization** – Installing persistence and ensuring access to the network is safe and stable.

4. **Privilege Escalation** – Gaining elevated privileges in the network.

5. **Action on Objectives** – Pivoting to relevant servers/hosts, exfiltrating data, installing additional malware, etc.

Given the gauntlet of products and configurations each network presents, it is important that offensive teams take steps to reduce exposure of their Intellectual Property (aka Tools, Tactics, and Procedures) at all phases of an attack. The cost of not doing so can be high – ask any team that has needed to re-roll their entire infrastructure, lost a useful technique to carelessness, or had to rewrite a piece of malware.

One important way to protect offensive IP is by preventing the detection of phishing payloads. Phishing is a common technique to gain initial access to organizations' networks. A typical phishing email will emulate correspondence from a trusted entity, with the aim of convincing the user to access a malicious web link or attachment. When clicked, the link or attachment will download a payload onto the user's system, ready for the user to execute. After execution, malware is deployed, giving access to the user's host and potentially compromising the security of an entire network. This is particularly dangerous when the user works for a large corporation or government entity that safeguards critical information.

To combat the rise in phishing emails that contain malicious documents, security vendors have integrated sandbox environments into their products. Because sandboxes provide a controlled environment for security analysts to observe malware, it is in the best interest of attackers to keep their malware from executing in a sandbox. To evade analysis, malicious payloads often contain checks against the properties of the host that would indicate whether or not the payload is being executed in a sandbox. If a host fails a check, the payload simply exits, or executes benign code. In this way, a payload might evade scrutiny from more skilled human analysts. Sandboxes also provide a pipeline of threat intelligence data. This data further helps defenders by providing clues about new trends in phishing techniques and malware authorship.

In this paper we present a novel sandbox detection method based on process list data. We compare the performance of two ML algorithms on this task: A Decision Tree classifier based on [3] and [2] and a two-layer Artificial Neural Network (ANN) implemented in Keras [5]. Empirical results show that both models perform well, and are both accurate enough to trust with automated malware deployment decisions. We conclude by highlighting several operational considerations that govern potential deployment of this technology in production settings, including Attribution, Sandbox drift, and Adversarial inputs.

### 1.1 Basic Sandbox Evasion

Successful execution of a payload on target largely depends on successful sandbox evasion. This can be accomplished via inline evasion techniques such as extended sleep times; logic that executes when specific conditions are met; or calls to trusted domains for special key-exchanges. However, many of these behaviors are well known and can be detected by sufficiently advanced sandboxes [1] [11]. Further, some sandbox checks are used so frequently in the context of malware, that the checks themselves become classified as malicious.

Common sandbox detection techniques include checking for recently used files, virtualization MAC addresses, the presence of a keyboard, domain membership, and so forth. However, in the escalating arms race of cybersecurity, security vendors clamp down on detectable information almost as quickly as malware developers begin to exploit it [4]. Moreover, as information for attackers grows harder to come by, the best checks are checks that provide additional information about a host, or the environment.

When attempting to gain initial access via phishing, one of the first pieces of information our team gathers from a host is a process list[3]. We manually review each process list in order to gather the security products installed on the host, the architecture, domain-joined status, and user context. The information is used to determine if it is safe to proceed with the next phase of the attack - installing persistence and deploying other tools. Additionally, a process list can be helpful when troubleshooting deployment failures: By knowing more about the execution environment, we can more confidently adjust our technique or payload for the next execution opportunity.

## 2 Related Work

To our knowledge, no previous research has explored the application of machine learning to sandbox detection via process list data. As it turns out, applications of ML to *any* form of offensive strategy are difficult to come by, primarily due to data scarcity within the execution environment and the lack of ability for offensive teams to share rich datasets. Accordingly, ML for penetration testing generally focuses on exploring already open-source data such as discovering vulnerabilities, for example by using a Naive Bayes classifier to identify the active web server [6] or by using a convolutional neural network to identify outdated (and hence likely vulnerable) web sites via visual inspection of screenshots [7]. ML has also been used for payload optimization by the penetration test tool DeepExploit [9], which examines system configurations and selects payloads via the asynchronous model A3C [10].

## 3 Methodology

In this research, our goal was to accurately classify sandboxes using the aforementioned process list data. A key advantage of this approach is its ability to generalize and automate sandbox checks, such that network operators aren't responsible for manually exiting malware that has been executed in a sandbox. Additionally, rather than gathering information from multiple sources on the host, thereby increasing activity on the target, simply collecting a process list is sufficient to gather the model's input data. How the information is gathered is not part of this paper, but common execution vectors include command execution, Win32 API calls, or shellcode injection. Each process list was posted to an external server for collection and processing. There were no sandbox detection checks used during data collection. Each process list was manually labelled as target (0), or sandbox (1).

[3] `tasklist.exe/ps`

### 3.1 Process List Data

A process list is a valuable piece of information. Not only is it a reflection of the security posture of the host, it is a reflection of the user, and a standard corporate image. Experience of the authors suggests that differences between "safe" (Appendix: Table 5) and "unsafe" (Appendix: Table 6) hosts are consistent across multiple organizations and sandboxes from multiple vendors, such that process list data could be generalized sufficiently to make an accurate classification.

Since both experience and empirical evidence (Appendix: Table 5, Table 6) suggest that process count and user context are major contributors to successful human classification[4], the following features were selected:

- Process Count
- User Count
- Process Count/User Count Ratio

**Process Count** - End user workstations typically have a lot of processes outside of the default Windows processes running: Multiple office products, security products, etc. Sandboxes just boot up, run the payload, and close. Sandboxes aren't using Excel, Word, Spotify, etc. all at the same time. Additionally, most end user workstations are Windows 10, while most sandboxes still run Windows 7. It's an important note, because the number of default processes on Windows 10 is significantly higher.

**User Count** – User count is a proxy for administrative privileges. On Windows, if the user is running in a medium-integrity context, the process list will only contain ownership information for processes the user owns (or has access to read). Otherwise, if the user is running in a high-integrity context, the process list will contain the owners for all processes. The difference in user count can be attributed to the fact that most sandboxes run payloads in a high-integrity context, but most organizations have removed administrative rights from their users.

**Process/User Ratio** - This feature is simply a combination of the previous features, and only seeks to support the inductive bias of the authors. Model accuracy was lower when this feature was omitted from training. Other features such as average process id, or a boolean indication of known safe processes running on the host would be alternative options not explored in our research.

[4] Even the casual observer will notice stark differences between each process list (Appendix: Table 5, Table 6)

| All Data (384 Samples) | | | | |
|---|---|---|---|---|
| **Feature** | **Min** | **Max** | **Mean** | **Std Dev** |
| Process Count | 9.0 | 305.0 | 80.0 | 59.9 |
| User Count | 1.0 | 17.0 | 2.5 | 1.7 |
| Ratio | 2.1 | 305.0 | 50.3 | 64.6 |
| **Safe (324 Samples)** | | | | |
| Process Count | 9.0 | 305.0 | 89.2 | 60.7 |
| User Count | 1.0 | 17.0 | 2.5 | 1.8 |
| Ratio | 2.1 | 305.0 | 57.2 | 68.0 |
| **Unsafe (60 Samples)** | | | | |
| Process Count | 11.0 | 56.0 | 30.6 | 11.65 |
| User Count | 1.0 | 4.0 | 2.9 | 1.0 |
| Ratio | 2.7 | 44.0 | 12.9 | 8.8 |

**Table 1.** Process list data collected via server uploads executed by phishing payloads. Environments were labeled "Safe" if a human analyst identified the process list as indicative of a production environment and "Unsafe" if the analyst classified it as a Sandbox environment.

### 3.2 Data Analysis and Limitations

An overview of the collected process lists and feature statistics is presented in Table 1. Of 384 collected process lists, 60 were judged to be sandbox environments and 324 were judged to be production environments. Process list length ranged from 9.0 - 305.0 active processes, with a mean length of 80.0 and a standard deviation of 59.9.

A number of observations can be made about the data depicted in Table 1. Firstly, user count is not a robust way of checking whether the payload was run as an administrator. While most sandboxes run payloads as an administrator, some do not, and this reflects in the user count due to the variance in Windows permissions.

Secondly, the max users in safe hosts is 17.0 (Table 1). It is unusual for a phishing target to have 17 users logged in, and is more indicative of a server of some sort, or would indicate a particular organization's remote access implementation. This data point could be confidently removed.

Finally, the data set is small by machine learning standards, and could affect the model's ability to generalize for successful classification. Additionally, the dataset was not cleaned or processed to remove oddities or potentially erroneous data points.

## 4 Algorithms

We applied two ML algorithms to our sandbox detection task: A Decision Tree Classifier and an Artificial Neural Network. We will see in Section 5 that both models performed well, but the Decision Tree performed best overall. These two algorithms were selected for their simplicity. It is of utmost importance that ML models deployed in offensive tasks are able to operate without much human oversight, particularly when many offensive teams lack data scientists.

### 4.1 Decision Trees

Decision Trees [3] are a non-parametric supervised learning method that can be easily visualized and interpreted, as seen in Figure 1. They are able to handle both numerical and categorical data, and do not require data normalization or other data preparation techniques. Potential drawbacks of Decision Trees include their susceptibility to overfitting and their difficulty representing XOR, parity, or multiplexor problems. They are also prone to overfitting in some contexts.

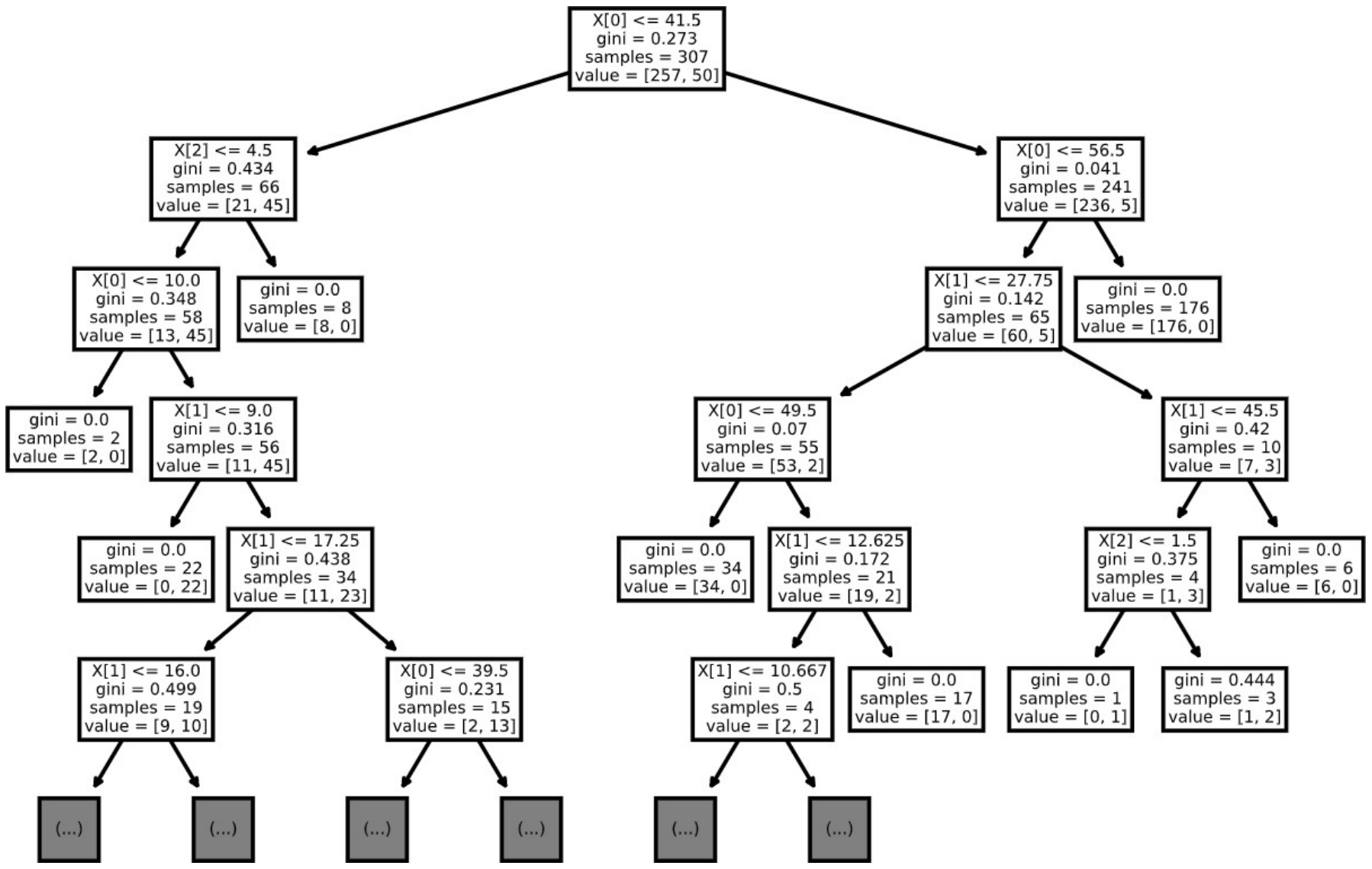

**Fig. 1.** Graphical depiction of a sandbox-detection Decision Tree Classifier generated using scikit-learn [12]. A Decision Tree learns to classify each input vector using if-then decision rules extracted from direct observations of the data.

Of the machine learning algorithms considered by our team, Decision Trees are closest to current human-driven methods of detection through a series of true/false checks. We trained our Decision tree using the features depicted in Table 1. Training was completed with a 80/20 split. No alterations were made to the raw features, such that the features of the process lists found in Appendix Table 5 and 6 were:

$$hosts = [(40, 4, 10), (220, 1, 220), \ldots]$$

We found the Decision Tree Classifier to be data efficient, quick to train, and effective at our sandbox detection task. Additionally, team members preferred the Decision Tree due to its implicit explainability. Further details can be found in Section 5 of this paper.

### 4.2 Artificial Neural Network

An Artificial Neural Network (ANN) is a biologically-inspired method of detecting predictable relationships between input samples and their associated training labels [13] [14]. ANNs can be difficult to train, but are able to represent complex functions and generalize well to previously unseen input configurations.

For sandbox detection, we used a 3 by 3 artificial neural network built with Keras using binary cross-entropy loss. Raw inputs were scaled with min-max, and a sigmoid activation function was used. The model was trained for 500 epochs, as depicted in Figure 2.

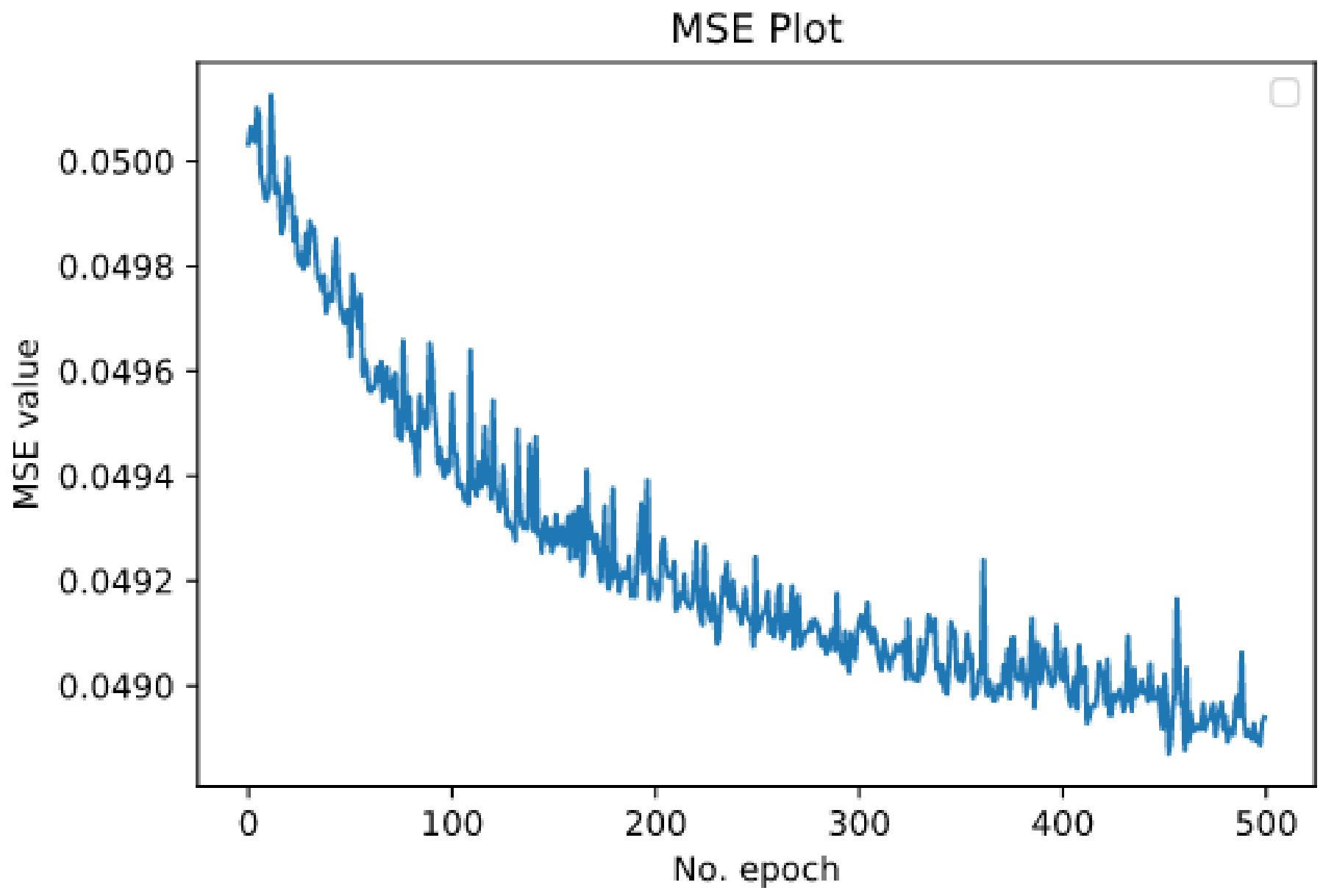

**Fig. 2.** ANN Mean Squared Error during training. Minimum loss was achieved within a few hundred learning cycles, not surprising given the relatively small size of the training data.

One challenge faced by our ANN was the fact that our dataset was both small and relatively "dirty". This is not unexpected given our use case, but it does create a challenge for a learning algorithm that usually requires thousands of training examples in order to generalize well. In an ideal scenario, we would have liked to collect a larger dataset, but real-world constraints of extracting process list information from initial access payloads in production environments made this infeasible.

## 5 Results

Results from our experiments are shown in Tables 2 and 3. The ANN achieved a classification accuracy of 92.71%. The Decision Tree Classifier was able to improve accuracy by 2.09% over the ANN, obtaining an overall classification accuracy of 94.80%.

| ANN Results | |
|---|---|
| Mean Absolute Error | 0.1188 |
| Mean Squared Error | 0.0573 |
| Accuracy | 92.71% |

**Table 2.** Sandbox detection accuracy for a two-layer network with clamped inputs and sigmoid activation function using binary cross-entropy loss. This model was able to generalize well to unseen data, identifying sandbox environments with high likelihood.

| Decision Tree Results | |
|---|---|
| True Positive | 66 |
| False Positive | 1 |
| False Negative | 3 |
| True Negative | 7 |
| Accuracy | 94.80% |

**Table 3.** Sandbox detection accuracy for a Decision Tree Classifier trained on the process list data depicted in Table 1. This model was able to outperform the ANN by 2.09%, and was preferred by our team because its decisions were transparent to humans.

Table 4 shows the calculated precision, recall, and F1-score for Safe and Unsafe execution environments respectively, along with macro and weighted averages accounting for the classification imbalance in the dataset. Examination of the data reveals that safe hosts had a high F1-score of 0.97, but the unsafe hosts had a lower than acceptable level of 0.78. This is likely due to the small sample size of unsafe hosts. From an operational security perspective, avoiding unsafe execution is far more important than achieving safe execution for the longevity of a piece of malware.

| Metrics | Precision | Recall | F1-score | support |
|---|---|---|---|---|
| Safe | 0.96 | 0.99 | 0.97 | 67 |
| Unsafe | 0.88 | 0.70 | 0.78 | 10 |
| Macro Average | 0.92 | 0.84 | 0.87 | 77 |
| Weighted Average | 0.95 | 0.95 | 0.95 | 77 |

**Table 4.** Decision Tree Metrics. Precision is the ratio of true positives to total predicted positives. Recall is the ratio of true positives to total actual positives. F1-score is defined as $2 * \frac{precision*recall}{precision+recall}$. A "Safe" label indicates a non-sandbox production environment in which malicious code can be safely deployed. An "Unsafe" label indicates a sandbox environment. Support indicates the number of samples contributing to the data calculations.

## 6 Discussion and Future Work

There are several operational factors that must be considered before deploying models to a production setting.

1. Attribution - If machine-learning models were embedded into malicious documents, these files would be easy to attribute to a particular group as this technique is not well known or widespread.
2. Sandbox drift - Overnight, all sandboxes could change, making all models and data irrelevant. The difference in Windows 10 and Windows 7 from a process count standpoint is significant. Or worse, sandboxes could change slowly, leading to inconsistent predictions.
3. Adversarial inputs - Depending on how payloads are subsequently staged, an analyst could submit false inputs to the dropper server, gaining access to payloads.
4. NLP techniques - tokenizing a process list rather than using a regex could be a more robust way of parsing process lists, as malware deployment methods will change. Additionally, NLP techniques such as document classification become possible.
5. Data Collection - A separate data collection effort would be ideal, such that production payloads would be separate from collection payloads. This would allow for data collection efforts to support offensive operations without interfering with production deployments. Additionally, once access to a network has been gained, any host in the network becomes a potential phishing target. Therefore, any user workstation process list gathered would be a legitimate data point even if it did not come from the initial access payload.

## 7 Further Research

Further research in this area should focus on the collection of a larger and more balanced dataset in order to improve classification accuracy. We would also like to explore methods for client-side classification. Because a trained model is comprised of static weights, they could be embedded into phishing payloads. This strategy, among others, should be examined for soundness and practicality.

Machine learning for offensive operations has typically been confined to *a priori* vulnerability analysis such as detecting specific server software or identifying outdated web sites. Our research demonstrates that machine learning can also be useful *in situ*. Even with limited training data, ML is able to allow the detection of sandbox environments with high accuracy, thus improving the likelihood that malicious payloads will remain hidden. Going forward, we hope to see more applications of machine learning for such purposes.

## 8 Conclusion

In this paper we have explored the use of machine learning for *in situ* offensive operations, and have shown that both a Decision Tree Classifier and an Artificial Neural Network are able to detect safe environments with high accuracy and with a strong F1-score, even given limited data. We have outlined several operational considerations that will affect the use of this technology in production settings, and have suggested promising avenues for future exploration. As part of this research, a tool called 'Deep-Drop' was developed as a machine-learning enabled dropper server. Deep-Drop contains all code and data mentioned in this paper.[5]

## References

1. Hira Agrawal, J. Alberi, L. Bahler, J. Micallef, A. Virodov, M. Magenheimer, S. Snyder, V. Debroy, and E. Wong. Detecting hidden logic bombs in critical infrastructure software. *7th International Conference on Information Warfare and Security, ICIW 2012*, pages 1 – 11, 01 2012.
2. Leo Breiman. Random forests. *Machine learning*, 45(1):5 – 32, 2001.
3. Leo Breiman, J. H. Friedman, R. A. Olshen, and C. J. Stone. *Classification and Regression Trees*. Statistics/Probability Series. Wadsworth Publishing Company, Belmont, California, U.S.A., 1984.
4. Alexander Chailytko and Stanislav Skuratovich. Defeating sandbox evasion: how to increase the successful emulation rate in your virtual environment, 2017.
5. François Chollet et al. Keras: Deep learning for humans. https://github.com/fchollet/keras, 2015.
6. Alisa Esage. Gyoithon: tool to make penetration testing with machine learning. https://www.securitynewspaper.com/2018/06/02/gyoithon-tool-make-penetration-testing-machine-learning/, 2018.
7. Bishop Fox. Eyeballer. https://github.com/bishopfox/eyeballer, 2019.
8. Eric M Hutchins, Michael J Cloppert, and Rohan M Amin. Intelligence-driven computer network defense informed by analysis of adversary campaigns and intrusion kill chains. *Leading Issues in Information Warfare & Security Research*, 1(1):80, 2011.
9. isao takaesu. Deepexploit. https://github.com/13o-bbr-bbq/machine¥_learning¥_security/tree/master/DeepExploit, 2019.

[5] https://github.com/moohax/Deep-Drop

10. Volodymyr Mnih, Adria Puigdomenech Badia, Mehdi Mirza, Alex Graves, Timothy Lillicrap, Tim Harley, David Silver, and Koray Kavukcuoglu. Asynchronous methods for deep reinforcement learning. In *International conference on machine learning*, pages 1928 – 1937, 2016.
11. Hassan Mourad. Sleeping your way out of the sandbox. https://www.sans.org/reading-room/whitepapers/malicious/sleeping-sandbox-35797, 2015.
12. F. Pedregosa, G. Varoquaux, A. Gramfort, V. Michel, B. Thirion, O. Grisel, M. Blondel, P. Prettenhofer, R. Weiss, V. Dubourg, J. Vanderplas, A. Passos, D. Cournapeau, M. Brucher, M. Perrot, and E. Duchesnay. Scikit-learn: Machine learning in Python. *Journal of Machine Learning Research*, 12:2825 – 2830, 2011.
13. Kevin L Priddy and Paul E Keller. *Artificial neural networks: an introduction*, volume 68. SPIE press, 2005.
14. Frank Rosenblatt. The perceptron: a probabilistic model for information storage and organization in the brain. *Psychological review*, 65(6):386, 1958.

# Appendix

| Safe Process List | | | | | |
|---|---|---|---|---|---|
| PID | PPID | ARCH | SESS | NAME | OWNER |
| 0 | 0 | | | System Process | |
| 4 | 0 | | | System | |
| 456 | 4 | | | smss.exe | |
| 596 | 536 | | | csrss.exe | |
| 784 | 536 | | | wininit.exe | |
| 792 | 776 | | | csrss.exe | |
| 864 | 784 | | | services.exe | |
| 872 | 784 | | | lsass.exe | |
| 1020 | 864 | | | svchost.exe | |
| 416 | 864 | | | svchost.exe | |
| 496 | 784 | | | fontdrvhost.exe | |
| 860 | 864 | | | svchost.exe | |
| 1032 | 864 | | | svchost.exe | |
| 1116 | 776 | | | winlogon.exe | |
| 1172 | 1116 | | | fontdrvhost.exe | |
| 1268 | 864 | | | svchost.exe | |
| 1316 | 864 | | | svchost.exe | |
| 1324 | 864 | | | svchost.exe | |
| 1332 | 864 | | | svchost.exe | |
| 1484 | 864 | | | svchost.exe | |
| 1496 | 864 | | | svchost.exe | |
| 1504 | 864 | | | svchost.exe | |
| 1588 | 864 | | | svchost.exe | |
| 1660 | 864 | | | svchost.exe | |
| 1712 | 864 | | | svchost.exe | |
| 1732 | 1116 | | | dwm.exe | |
| 1788 | 864 | | | svchost.exe | |
| 1900 | 864 | | | svchost.exe | |
| 1912 | 864 | | | svchost.exe | |
| 2032 | 864 | | | svchost.exe | |
| 1168 | 864 | | | svchost.exe | |
| 2072 | 864 | | | nvvsvc.exe | |
| 2080 | 864 | | | nvscpapisvr.exe | |
| 2180 | 864 | | | svchost.exe | |
| 2216 | 2072 | | | nvxdsync.exe | |
| 2296 | 864 | | | svchost.exe | |
| 2328 | 864 | | | svchost.exe | |
| 2336 | 864 | | | svchost.exe | |
| 2364 | 864 | | | svchost.exe | |
| 2384 | 864 | | | svchost.exe | |
| 2460 | 864 | | | svchost.exe | |
| 2512 | 864 | | | svchost.exe | |
| 2540 | 864 | | | svchost.exe | |
| 2552 | 1268 | | | WUDFHost.exe | |
| 2580 | 1168 | | | dasHost.exe | |
| 2676 | 864 | | | svchost.exe | |
| 2684 | 864 | | | svchost.exe | |
| 2732 | 864 | | | svchost.exe | |
| 2940 | 864 | | | svchost.exe | |
| 3004 | 864 | | | svchost.exe | |
| 2472 | 864 | | | svchost.exe | |
| 3068 | 864 | | | igfxCUIService.exe | |
| 3092 | 864 | | | svchost.exe | |
| 3180 | 864 | | | svchost.exe | |
| 3192 | 864 | | | svchost.exe | |
| 3544 | 1268 | | | WUDFHost.exe | |
| 3656 | 864 | | | svchost.exe | |
| 3820 | 864 | | | svchost.exe | |
| 3876 | 864 | | | RtkAudioService64.exe | |
| 3888 | 864 | | | SavService.exe | |
| 4000 | 864 | | | svchost.exe | |
| 3908 | 864 | | | SearchIndexer.exe | |
| 4124 | 3876 | | | RAVBg64.exe | |
| 4148 | 864 | | | svchost.exe | |
| 4156 | 864 | | | svchost.exe | |
| 4164 | 864 | | | PulseSecureService.exe | |
| 4200 | 864 | | | svchost.exe | |
| 4308 | 864 | | | svchost.exe | |
| 4428 | 864 | | | svchost.exe | |
| 4532 | 864 | | | svchost.exe | |
| 4620 | 864 | | | SCFManager.exe | |
| 4636 | 864 | | | spoolsv.exe | |
| 4892 | 864 | | | SACSRV.exe | |
| 4952 | 864 | | | SCFService.exe | |
| 5160 | 864 | | | mDNSResponder.exe | |
| 5172 | 864 | | | armsvc.exe | |
| 5184 | 864 | | | OfficeClickToRun.exe | |
| 5196 | 864 | | | svchost.exe | |
| 5236 | 864 | | | AppleMobileDeviceService.exe | |
| 5260 | 864 | | | svchost.exe | |
| 5268 | 864 | | | IntelCpHDCPSvc.exe | |
| 5276 | 864 | | | AdminService.exe | |
| 5284 | 864 | | | LogiRegistryService.exe | |
| 5324 | 864 | | | svchost.exe | |
| 5340 | 864 | | | esif_uf.exe | |
| 5352 | 864 | | | svchost.exe | |
| 5360 | 864 | | | FoxitConnectedPDFService.exe | |
| 5400 | 864 | | | ALsvc.exe | |
| 5412 | 864 | | | svchost.exe | |
| 5420 | 864 | | | swc service.exe | |
| 5448 | 864 | | | ManagementAgentNT.exe | |
| 5480 | 864 | | | SecurityHealthService.exe | |
| 5492 | 864 | | | RouterNT.exe | |
| 5600 | 864 | | | SAVAdminService.exe | |
| 5648 | 864 | | | swi service.exe | |
| 5692 | 864 | | | SntpService.exe | |
| 5708 | 864 | | | svchost.exe | |
| 5720 | 864 | | | sqlwriter.exe | |
| 5732 | 864 | | | ssp.exe | |
| 5760 | 864 | | | swi filter.exe | |
| 5776 | 864 | | | TBear.Maintenance.exe | |
| 5788 | 864 | | | WavesSysSvc64.exe | |
| 5796 | 864 | | | TeamViewer Service.exe | |
| 5804 | 864 | | | svchost.exe | |
| 5812 | 864 | | | svchost.exe | |
| 5832 | 864 | | | svchost.exe | |
| 5944 | 864 | | | svchost.exe | |
| 5952 | 864 | | | svchost.exe | |
| 6132 | 5760 | | | swi_fc.exe | |
| 6432 | 4 | | | Memory Compression | |
| 6716 | 864 | | | svchost.exe | |
| 7152 | 864 | | | IntelCpHeciSvc.exe | |
| 7220 | 4164 | | | PulseSecureService.exe | |
| 8764 | 864 | | | sdcservice.exe | |
| 9076 | 864 | | | svchost.exe | |
| 5316 | 864 | | | svchost.exe | |
| 9084 | 864 | | | csia.exe | |
| 3636 | 864 | | | svchost.exe | |
| 3148 | 864 | | | svchost.exe | |
| 4688 | 5340 | x64 | 1 | esif_assist_64.exe | CORP\¡REDACTED¿ |
| 5728 | 2032 | x64 | 1 | sihost.exe | CORP\¡REDACTED¿ |
| 8020 | 864 | x64 | 1 | svchost.exe | CORP\¡REDACTED¿ |
| 7968 | 864 | x64 | 1 | svchost.exe | CORP\¡REDACTED¿ |
| 6448 | 1660 | x64 | 1 | itype.exe | CORP\¡REDACTED¿ |
| 7488 | 1660 | x64 | 1 | ipoint.exe | CORP\¡REDACTED¿ |
| 1412 | 1660 | x64 | 1 | taskhostw.exe | CORP\¡REDACTED¿ |
| 6212 | 864 | | | svchost.exe | |
| 3848 | 864 | | | PresentationFontCache.exe | |
| 1048 | 5884 | x64 | 1 | explorer.exe | CORP\¡REDACTED¿ |
| 9236 | 3520 | x64 | 1 | igfxEM.exe | CORP\¡REDACTED¿ |
| 9532 | 416 | x64 | 1 | ShellExperienceHost.exe | CORP\¡REDACTED¿ |
| 9768 | 416 | x64 | 1 | RuntimeBroker.exe | CORP\¡REDACTED¿ |
| 10132 | 1900 | x64 | 1 | TabTip.exe | |
| 10192 | 10132 | x86 | 1 | TabTip32.exe | |
| 10244 | 864 | | | svchost.exe | |
| 11596 | 10396 | x64 | 1 | chrome.exe | CORP\¡REDACTED¿ |
| 9996 | 11596 | x64 | 1 | chrome.exe | CORP\¡REDACTED¿ |
| 10744 | 11596 | x64 | 1 | chrome.exe | CORP\¡REDACTED¿ |
| 12316 | 416 | x64 | 1 | SystemSettingsBroker.exe | CORP\¡REDACTED¿ |
| 12464 | 864 | | | svchost.exe | |
| 12624 | 416 | | | WmiPrvSE.exe | |
| 13176 | 11596 | x64 | 1 | chrome.exe | CORP\¡REDACTED¿ |
| 13208 | 11596 | x64 | 1 | chrome.exe | CORP\¡REDACTED¿ |
| 13224 | 11596 | x64 | 1 | chrome.exe | CORP\¡REDACTED¿ |
| 13236 | 11596 | x64 | 1 | chrome.exe | CORP\¡REDACTED¿ |
| 2776 | 11596 | x64 | 1 | chrome.exe | CORP\¡REDACTED¿ |
| 780 | 11596 | x64 | 1 | chrome.exe | CORP\¡REDACTED¿ |
| 1544 | 11596 | x64 | 1 | chrome.exe | CORP\¡REDACTED¿ |
| 13332 | 11596 | x64 | 1 | chrome.exe | CORP\¡REDACTED¿ |
| 15048 | 1048 | x64 | 1 | MSASCuiL.exe | CORP\¡REDACTED¿ |
| 15224 | 1048 | x64 | 1 | RtkNGUI64.exe | CORP\¡REDACTED¿ |
| 15332 | 1048 | x64 | 1 | RAVBg64.exe | CORP\¡REDACTED¿ |
| 14540 | 1048 | x64 | 1 | WavesSvc64.exe | CORP\¡REDACTED¿ |
| 800 | 1048 | x64 | 1 | SACMonitor.exe | CORP\¡REDACTED¿ |
| 14648 | 1048 | x64 | 1 | LCore.exe | CORP\¡REDACTED¿ |
| 13152 | 1048 | x64 | 1 | RtkUGui64.exe | CORP\¡REDACTED¿ |
| 15444 | 1048 | x64 | 1 | iTunesHelper.exe | CORP\¡REDACTED¿ |
| 16144 | 1048 | x64 | 1 | DellSystemDetect.exe | CORP\¡REDACTED¿ |
| 9116 | 1048 | x86 | 1 | ONENOTEM.EXE | CORP\¡REDACTED¿ |
| 15840 | 10652 | x86 | 1 | Pulse.exe | CORP\¡REDACTED¿ |
| 16316 | 864 | | | iPodService.exe | |
| 16124 | 10652 | x86 | 1 | ALMon.exe | CORP\¡REDACTED¿ |
| 15980 | 10652 | x86 | 1 | jusched.exe | CORP\¡REDACTED¿ |
| 15928 | 1660 | x64 | 1 | RAVBg64.exe | CORP\¡REDACTED¿ |
| 14760 | 416 | x64 | 1 | unsecapp.exe | CORP\¡REDACTED¿ |
| 15412 | 864 | | | svchost.exe | |
| 13648 | 1048 | x64 | 1 | Slack.exe | CORP\¡REDACTED¿ |
| 9932 | 13648 | x64 | 1 | Slack.exe | CORP\¡REDACTED¿ |
| 13656 | 13648 | x64 | 1 | Slack.exe | CORP\¡REDACTED¿ |
| 14700 | 1048 | x86 | 1 | OUTLOOK.EXE | CORP\¡REDACTED¿ |
| 14416 | 13648 | x64 | 1 | Slack.exe | CORP\¡REDACTED¿ |
| 14428 | 416 | x64 | 1 | iexplore.exe | CORP\¡REDACTED¿ |
| 15308 | 14428 | x86 | 1 | iexplore.exe | CORP\¡REDACTED¿ |
| 9404 | 864 | x64 | 1 | svchost.exe | CORP\¡REDACTED¿ |
| 12400 | 416 | x64 | 1 | dllhost.exe | CORP\¡REDACTED¿ |
| 11324 | 864 | | | svchost.exe | |
| 16096 | 864 | | | svchost.exe | |
| 11140 | 416 | | | dllhost.exe | |
| 2964 | 11596 | x64 | 1 | chrome.exe | CORP\¡REDACTED¿ |
| 13392 | 1048 | x86 | 1 | HprSnap6.exe | CORP\¡REDACTED¿ |
| 9844 | 13392 | x64 | 1 | TsHelper64.exe | CORP\¡REDACTED¿ |
| 14880 | 864 | | | svchost.exe | |
| 2980 | 13648 | x64 | 1 | Slack.exe | CORP\¡REDACTED¿ |
| 1152 | 11596 | x64 | 1 | chrome.exe | CORP\¡REDACTED¿ |
| 8964 | 416 | x64 | 1 | CertEnrollCtrl.exe | CORP\¡REDACTED¿ |
| 15488 | 11596 | x64 | 1 | chrome.exe | CORP\¡REDACTED¿ |
| 16836 | 864 | | | svchost.exe | |
| 4616 | 1268 | | | WUDFHost.exe | |
| 9416 | 416 | x64 | 1 | InstallAgent.exe | CORP \¡REDACTED¿ |
| 7336 | 416 | x64 | 1 | InstallAgentUserBroker.exe | CORP \¡REDACTED¿ |
| 11724 | 864 | | | svchost.exe | |
| 17420 | 864 | | | svchost.exe | |
| 17364 | 11596 | x64 | 1 | chrome.exe | CORP\¡REDACTED¿ |
| 15372 | 11596 | x64 | 1 | chrome.exe | CORP\¡REDACTED¿ |
| 10752 | 864 | | | svchost.exe | |
| 5672 | 11596 | x64 | 1 | chrome.exe | CORP\¡REDACTED¿ |
| 18392 | 416 | x64 | 1 | Microsoft.StickyNotes.exe | CORP\¡REDACTED¿ |
| 476 | 416 | x64 | 1 | SkypeHost.exe | CORP\¡REDACTED¿ |
| 11656 | 11596 | x64 | 1 | chrome.exe | CORP\¡REDACTED¿ |
| 6672 | 11596 | x64 | 1 | chrome.exe | CORP\¡REDACTED¿ |
| 13044 | 416 | x86 | 1 | OneDrive.exe | CORP\¡REDACTED¿ |
| 9920 | 416 | x64 | 1 | SearchUI.exe | CORP\¡REDACTED¿ |
| 12768 | 11596 | x64 | 1 | chrome.exe | CORP\¡REDACTED¿ |
| 1968 | 11596 | x64 | 1 | chrome.exe | CORP\¡REDACTED¿ |
| 10004 | 11596 | x64 | 1 | chrome.exe | CORP\¡REDACTED¿ |
| 13716 | 11596 | x64 | 1 | chrome.exe | CORP\¡REDACTED¿ |
| 17216 | 11596 | x64 | 1 | chrome.exe | CORP\¡REDACTED¿ |
| 14400 | 11596 | x64 | 1 | chrome.exe | CORP\¡REDACTED¿ |
| 12348 | 11596 | x64 | 1 | chrome.exe | CORP\¡REDACTED¿ |
| 10496 | 1116 | | | LogonUI.exe | |
| 16428 | 416 | x64 | 1 | LockAppHost.exe | CORP\¡REDACTED¿ |
| 1632 | 416 | x64 | 1 | LockApp.exe | CORP\¡REDACTED¿ |
| 16156 | 416 | x64 | 1 | ApplicationFrameHost.exe | CORP\¡REDACTED¿ |
| 16540 | 416 | x64 | 1 | SystemSettings.exe | CORP\¡REDACTED¿ |
| 13948 | 416 | x64 | 1 | Calculator.exe | CORP\¡REDACTED¿ |
| 11752 | 416 | x64 | 1 | Microsoft.Photos.exe | CORP\¡REDACTED¿ |
| 16568 | 864 | | | svchost.exe | |
| 11200 | 3592 | | | ¡REDACTED¿.exe | |
| 5684 | 3820 | x64 | 0 | audiodg.exe | |
| 16856 | 1660 | x64 | 1 | ¡REDACTED¿.exe | CORP\¡REDACTED¿ |
| 6416 | 16856 | x64 | 1 | conhost.exe | CORP\¡REDACTED¿ |

Table 5: Safe Process List

| Unsafe Process List | | | | | |
|---|---|---|---|---|---|
| PID | PPID | ARCH | SESS | NAME | OWNER |
| 0 | 0 | | | System Process | |
| 4 | 0 | x86 | 0 | System | |
| 244 | 4 | x86 | 0 | smss.exe | NT AUTHORITY¥SYSTEM |
| 352 | 304 | x86 | 0 | csrss.exe | NT AUTHORITY¥SYSTEM |
| 400 | 304 | x86 | 0 | wininit.exe | NT AUTHORITY¥SYSTEM |
| 408 | 392 | x86 | 1 | csrss.exe | NT AUTHORITY¥SYSTEM |
| 440 | 392 | x86 | 1 | winlogon.exe | NT AUTHORITY¥SYSTEM |
| 504 | 400 | x86 | 0 | services.exe | NT AUTHORITY¥SYSTEM |
| 512 | 400 | x86 | 0 | lsass.exe | NT AUTHORITY¥SYSTEM |
| 520 | 400 | x86 | 0 | lsm.exe | NT AUTHORITY¥SYSTEM |
| 628 | 504 | x86 | 0 | svchost.exe | NT AUTHORITY¥SYSTEM |
| 696 | 504 | x86 | 0 | svchost.exe | NT AUTHORITY¥NETWORK SERVICE |
| 744 | 504 | x86 | 0 | svchost.exe | NT AUTHORITY¥LOCAL SERVICE |
| 864 | 504 | x86 | 0 | svchost.exe | NT AUTHORITY¥SYSTEM |
| 932 | 504 | x86 | 0 | svchost.exe | NT AUTHORITY¥LOCAL SERVICE |
| 972 | 504 | x86 | 0 | svchost.exe | NT AUTHORITY¥SYSTEM |
| 1140 | 504 | x86 | 0 | svchost.exe | NT AUTHORITY¥NETWORK SERVICE |
| 1344 | 504 | x86 | 0 | spoolsv.exe | NT AUTHORITY¥SYSTEM |
| 1380 | 504 | x86 | 1 | taskhost.exe | bea-chi-t-7pr01¥John Doe |
| 1408 | 504 | x86 | 0 | svchost.exe | NT AUTHORITY¥LOCAL SERVICE |
| 1512 | 972 | x86 | 1 | taskeng.exe | bea-chi-t-7pr01¥John Doe |
| 1548 | 504 | x86 | 0 | mfemms.exe | NT AUTHORITY¥SYSTEM |
| 1636 | 1548 | x86 | 0 | mfevtps.exe | NT AUTHORITY¥SYSTEM |
| 1696 | 1512 | x86 | 1 | cmd.exe | bea-chi-t-7pr01 ¥John Doe |
| 1708 | 1548 | x86 | 0 | mfehcs.exe | NT AUTHORITY ¥SYSTEM |
| 1940 | 504 | x86 | 0 | sppsvc.exe | NT AUTHORITY ¥NETWORK SERVICE |
| 2016 | 504 | x86 | 0 | svchost.exe | NT AUTHORITY ¥NETWORK SERVICE |
| 340 | 408 | x86 | 1 | conhost.exe | bea-chi-t-7pr01 ¥John Doe |
| 256 | 1696 | x86 | 1 | cmd.exe | bea-chi-t-7pr01 ¥John Doe |
| 308 | 256 | x86 | 1 | GoatCasper.exe | bea-chi-t-7pr01 ¥John Doe |
| 1780 | 864 | x86 | 1 | dwm.exe | bea-chi-t-7pr01 ¥John Doe |
| 1008 | 1748 | x86 | 1 | explorer.exe | bea-chi-t-7pr01 ¥John Doe |
| 1436 | 1008 | x86 | 1 | jusched.exe | bea-chi-t-7pr01 ¥John Doe |
| 264 | 504 | x86 | 0 | svchost.exe | NT AUTHORITY ¥LOCAL SERVICE |
| 648 | 504 | x86 | 0 | svchost.exe | NT AUTHORITY ¥SYSTEM |
| 220 | 504 | x86 | 0 | SearchIndexer. | NT AUTHORITY ¥SYSTEM |
| 2328 | 1008 | x86 | 1 | ¡REDACTED¿ | bea-chi-t-7pr01 ¥John Doe |
| 2700 | 628 | x86 | 1 | Setup.exe | bea-chi-t-7pr01 ¥John Doe |
| 2240 | 504 | x86 | 0 | msiexec.exe | NT AUTHORITY¥SYSTEM |
| 3272 | 2240 | x86 | 1 | msiexec.exe | bea-chi-t-7pr01¥John Doe |
| 3056 | 768 | x86 | 0 | MpCmdRun.exe | NT AUTHORITY ¥NETWORK SERVICE |

**Table 6.** Unsafe Process List